\documentclass[aps,pra,reprint,groupedaddress,showkeys,showpacs]{revtex4-2}
\usepackage{graphicx}
\usepackage{dcolumn}
\usepackage{bm}
\usepackage[colorlinks,linkcolor=blue,anchorcolor=blue,citecolor=blue,urlcolor=blue]{hyperref}
\usepackage{amsmath}
\begin{document}


\title{Photon blockade effect from synergistic optical parametric amplification and driving force in Kerr-medium single-mode cavity}


\author{ZHANG Zhiqiang}
\email{zhangzhiqiang08@gmail.com}

\thanks{\\\textcolor[rgb]{1.00,0.00,0.00}{This paper is an English translated version of the original Chinese paper published in Acta Physica Sinica. Please cite the paper as: ZHANG Zhiqiang, Photon blockade effect from synergistic optical parametric amplification and driving force in Kerr-medium single-mode cavity. Acta Phys. Sin., 2025, 74(16): 164205. DOI: \href{https://doi.org/10.7498/aps.74.20250712} {10.7498/aps.74.20250712}}}



\affiliation{Department of College Physics, School of Intelligent Construction, Zhengzhou Business University, Zhengzhou 451200, Henan, China}


\date{\today}

\begin{abstract}
By combining analytical solutions and numerical simulations, we investigate the control mechanism of photon blockade effects in a hybrid quantum system consisting of a Kerr-medium single-mode cavity coupled with an optical parametric amplifier (OPA). To study photon blockade in the system, the dynamics are described by a master equation derived from the effective Hamiltonian, which considers single-mode cavity decay. In order to obtain analytical solutions under optimal photon blockade conditions, the quantum state of the system is expanded to the two-photon level based on the Fock state, and the steady-state probability amplitudes are derived by solving the Schrödinger equation, thereby yielding analytical expressions for the optimal photon blockade regime. The results demonstrate that photon blockade can be achieved in the system at appropriate parameters. Comparative analysis shows excellent agreement between the analytical results and numerical simulations of the equal-time second-order correlation function, validating both the correctness of the analytical solutions and the effectiveness of photon blockade in the system. The numerical results show that the average photon number significantly increases under resonant conditions, providing theoretical support for optimizing single-photon source brightness, which is essential for achieving high-brightness single-photon sources. Furthermore, variations in the driving phase can cause the optimal photon blockade region to shift in the two-dimensional parameter space of driving strength and OPA nonlinear coefficient, and even reverse the opening direction of the parabolic-shaped optimal blockade region. Both numerical and theoretical results confirm the regulatory effect of the driving phase on photon blockade. Additionally, the influence of Kerr nonlinearity is examined. The results show that photon blockade persists robustly over a broad range of Kerr nonlinear strengths, exhibiting universal characteristics. Physical mechanism analysis indicates that the photon blockade effect originates from destructive quantum interference between two photon transition pathways in the system under specific parameters, effectively suppressing two-photon excitation. Although Kerr nonlinearity modulates the energy levels of the system, it does not affect the quantum interference pathways, thus keeping the photon blocking effect stable over a wide parameter range.
\end{abstract}



\maketitle

\section{Introduction}

At present, quantum information technology is in the ascendant. As the core device of quantum information technology, the performance of single photon light source directly determines the feasibility of frontier applications such as quantum computing, quantum communication and quantum precision measurement. Among many single-photon generation mechanisms, single-photon sources based on photon blockade effect\cite{ref1,ref2} have become a hot research topic because of their significant advantages such as high purity, high repetition rate and integration. In 2005, Birnbaum et al realized the photon blockade effect experimentally for the first time\cite{ref3} by using a single-atom strong coupling optical cavity system, which laid the foundation for the development of single-photon source. In 2018, Snijders \emph{et al.}\cite{ref4} first observed the unconventional photon blockade effect experimentally by constructing a dual-mode optical cavity and using the asymmetric coupling between the orthogonal polarization mode and the quantum dot, and the single photon emission rate was one order of magnitude higher than that of the traditional scheme. At the same time, Vaneph et al designed a double-coupled superconducting cavity structure\cite{ref5} to achieve photon blockade effect in the microwave band by controlling the nonlinearity of the cavity mode. In 2025, Ding et al developed a single-photon source based on a quantum dot embedded in a tunable microcavity\cite{ref6}. By optimizing the design of the optical microcavity and pulse shaping technology, the overall system efficiency of the single-photon source reached 71.2\%, the photon indistinguishability was as high as 98.56\%, and the multi-photon error rate was as low as 2.05\%, showing extremely high single-photon purity.

Photon blockade is a non-classical effect in quantum optics, which means that under certain conditions, the existence of a single photon will inhibit the transmission or generation of subsequent photons, resulting in the output light field showing sub-Poissonian statistics and antibunching characteristics. This effect can be quantitatively characterized by the second-order correlation function. When the value of the second-order correlation function is less than 1, the existence of photon blockade in the system can be confirmed. Photon blockade can be divided into two types according to its physical mechanism: conventional photon blockade (CPB)\cite{ref7,ref8,ref9,ref10} and unconventional photon blockade (UPB)\cite{ref1,ref11,ref12,ref13}. CPB originates from the anharmonic splitting of energy levels caused by strong nonlinear interaction, which makes the frequency mismatch between single-photon resonant excitation and multi-photon transition, thus realizing single-photon selective excitation. UPB is a nonlinear quantum optical effect based on quantum interference cancellation of multi-photon transition paths. Its physical essence is that different quantum paths in the system satisfy interference cancellation under specific parameters, selectively suppressing the population of two-photon states while maintaining single-photon emission. Unlike CPB, UPB can be implemented under weakly nonlinear conditions.

In recent years, nonlinear optical cavity system provides an ideal research platform for the realization of controllable photon blockade effect. Many scholars have made extensive research on single-mode optical cavity, double-mode optical cavity, multiple optical cavities, the interaction between optical cavity and atom, and photon blockade in nonreciprocal optical system, and have made many progresses in the field of photon blockade. For a single-mode optical cavity system, the photon blockade effect can be generated by introducing physical mechanisms such as Kerr nonlinear\cite{ref14,ref15} or optical parametric amplification (OPA)\cite{ref16}. In addition, by introducing the frequency degree of freedom of photons as a new control dimension, the researchers proposed a novel photon blockade mechanism based on the frequency response characteristics of nonlinear optical cavities: the efficient photon blockade effect is realized by using the non-uniform quantum response of the cavity to driving fields with different frequencies\cite{ref17}. The study of photon blockade in a two-mode cavity system is mainly focused on the interaction between the two cavity modes and how to achieve photon blockade through this interaction\cite{ref18,ref19,ref20}. For example, in a two-mode cavity optomechanical system, strong photon antibunching can be achieved by adjusting the coupling strength between the two cavity modes and the driving condition\cite{ref20}. In addition, it is found that the two-mode cavity system can achieve unconventional photon blockade through quantum interference between different paths. This mechanism overcomes the limitation of conventional photon blockade, which typically requires large nonlinear strength\cite{ref21,ref22}. The interaction between multiple optical cavities provides more possibilities to realize the complex photon blockade effect. By coupling multiple cavities together, a system with a specific energy level structure can be formed, thus realizing photon blockade\cite{ref23,ref24}. In the cavity-atom coupling system, the coupling between the atom and the cavity mode can produce nonlinear effects, thus realizing photon blockade. The photon blockade effect can be effectively enhanced by adjusting the coupling strength between the atom and the cavity mode\cite{ref25,ref26} and the energy level structure of the atom\cite{ref27}. In addition, it is shown that a tunable unconventional photon blockade can be achieved by introducing a phase-controllable driving field into the single-atom-cavity system\cite{ref28,ref29}. The latest theoretical research is further extended to the nonreciprocal system\cite{ref30,ref31,ref32,ref33,ref34,ref35,ref36,ref37,ref38}, which introduces controllable nonreciprocity into the multipath quantum system to achieve one-way blockade of photon transmission and selective manipulation of quantum states. The results theoretically predict that a significant nonreciprocal photon blockade phenomenon\cite{ref39} can be observed in non-Hermitian cavities, and demonstrate the possibility of realizing robust photon blockade based on topological protection mechanism\cite{ref40,ref41}. At present, the theoretical research frontier of photon blockade is developing in the directions of multi-physical field cooperative control, non-Hermitian quantum optical effect and non-reciprocal quantum optics, which expands the theoretical research scope and application boundary of photon blockade.

In this paper, the quantum system composed of a Kerr-medium single-mode cavity and an optical parametric amplifier (OPA) is considered. The control mechanism of the photon blockade effect by the nonlinear coefficient of the OPA, the strength of the driving force, the phase of the driving force and the strength of the Kerr nonlinearity is studied by combining numerical analytical solution with numerical simulation. The results show that both the analytical and numerical results confirm that photon blockade can exist in the system under appropriate parameters, and the average photon number increases significantly when the system is in resonance, and the phase of the driving force has a significant regulatory effect on the photon blockade effect in the system. Furthermore, in a wide range of the Kerr nonlinearity strength, the system always exhibits a significant photon blockade effect, showing a typical universal photon blockade feature. Finally, the energy level structure and photon transition path are systematically analyzed to reveal the physical mechanism of the photon blockade effect.

\section{Physical model}
The system consists of a single-mode cavity and an optical parametric amplifier (OPA), with a Kerr medium present inside the cavity. Due to the existence of Kerr nonlinearity and OPA nonlinearity in the system, the Hamiltonian of the system is\cite{ref16,ref42,ref43}.

\begin{equation}\label{eqn1}
  {\hat H_{\mathrm{G}}} = {\omega _a}{\hat a^\dagger }\hat a + U{\hat a^\dagger }{\hat a^\dagger }\hat a\hat a + {\text{i}}G\left( {{{\hat a}^\dagger }{{\hat a}^\dagger } - \hat a\hat a} \right),
\end{equation}
where $\omega _a$ is the eigenfrequency of the single-mode cavity, $\hat a^\dagger $ and $\hat a$ are the photon creation and annihilation operators of the single-mode cavity, respectively, $U$ is the Kerr nonlinearity strength; $G$ is the optical parametric amplifier nonlinearity coefficient. The external driving force on the cavity is of the form
\begin{equation}\label{eqn2}
  {\hat H_{\text{d}}} = F\left( {{{\hat a}^\dagger }{{\text{e}}^{{\text{i}}\phi }}{{\text{e}}^{ - {\text{i}}{\omega _{\mathrm{d}}}t}} + \hat a{{\text{e}}^{ - {\text{i}}\phi }}{{\text{e}}^{{\text{i}}{\omega _{\mathrm{d}}}t}}} \right),
\end{equation}
where $F$ is the strength of the driving force, $\phi$ is the phase of the driving force, and $\omega_d$ is the frequency of the driving forces.
In order to study the evolution of the system, a rotating frame with respect to the frequency of the control field is adopted. Defining a rotating operator $\hat R = \exp \left( {{\text{i}}{\omega _{\text{d}}}t{{\hat a}^\dagger }\hat a} \right)$, the effective form of the total Hamiltonian $\hat H = {\hat H_{\text{G}}} + {\hat H_{\text{d}}}$ can be expressed as

\begin{align}\label{eqn3}
\begin{split} 
{\hat H_{{\text{eff}}}} =\;& \varDelta {\hat a^\dagger }\hat a + U{\hat a^\dagger }{\hat a^\dagger }\hat a\hat a + {\text{i}}G\left( {{{\hat a}^\dagger }{{\hat a}^\dagger } - \hat a\hat a} \right) \\
&+ F\left( {{{\hat a}^\dagger }{{\text{e}}^{{\text{i}}\phi }} + \hat a{{\text{e}}^{ - {\text{i}}\phi }}} \right),
\end{split}
\end{align}
where $\varDelta = {\omega _{\text{a}}} - {\omega _{\text{d}}}$ is the detuning of the eigenfrequency of the single-mode cavity from the frequency of the driving force.

To investigate photon blockade in a single-mode cavity, the dynamics of the system under photon decay can be described by the photon blockade master equation, which takes the following form. 

\begin{equation}\label{eqn4}
  \frac{{\partial \hat \rho }}{{\partial t}} = {\text{i}}\left[ {{{\hat H}_{{\text{eff}}}},\hat \rho } \right] + \frac{\kappa }{2}\left( {2{{\hat a}^\dagger }\hat \rho \hat a - {{\hat a}^\dagger }\hat a\hat \rho - \hat \rho {{\hat a}^\dagger }\hat a} \right),
\end{equation}
where $\rho$ is the quantum state and $\kappa$ is the photon decay rate of the single-mode cavity.
In theoretical studies, researchers usually use the equal-time second-order correlation function ${g^{\left( 2 \right)}}\left( 0 \right)$ to describe the steady-state statistical characteristics of photons in the system, and the equal-time second-order correlation function ${g^{\left( 2 \right)}}\left( 0 \right)$ is defined as follows:
\begin{equation}\label{eqn5}
{g^{\left( 2 \right)}}\left( 0 \right) = \frac{{\left\langle {{{\hat a}^\dagger }{{\hat a}^\dagger }\hat a\hat a} \right\rangle }}{{{{\left\langle {{{\hat a}^\dagger }\hat a} \right\rangle }^2}}} = \frac{{{\text{Tr}}\left( {{\rho _{\text{s}}}{{\hat a}^\dagger }{{\hat a}^\dagger }\hat a\hat a} \right)}}{{{{\left[ {{\text{Tr}}\left( {{\rho _{\text{s}}}{{\hat a}^\dagger }\hat a} \right)} \right]}^2}}},
\end{equation}
here, $\text{Tr}( )$ denotes taking the trace of the matrix, and $\rho _\text{s}$ is the steady state of the system. When the equal-time second-order correlation function of the system satisfies ${g^{\left( 2 \right)}}\left( 0 \right)<1$, the photon number distribution of the system is in a sub-Poissonian state, indicating that photon blockade occurs in the system.

In the study of single-photon sources, brightness is a key performance metric, defined as the average photon number of the system:
\begin{equation}\label{eqn6}
N = \left\langle {{{\hat a}^\dagger }\hat a} \right\rangle = {\text{Tr}}\left( {{\rho _{\text{s}}}{{\hat a}^\dagger }\hat a} \right).
\end{equation}
The average photon number directly reflects the efficiency of a single photon source to emit available single photons per unit time, and is an important parameter to measure the practicability of a single photon source.

For a system with a known Hamiltonian, the equal-time second-order correlation function ${g^{\left( 2 \right)}}\left( 0 \right)$and the average photon number $N$ can be obtained from the master equation by numerical calculation or theoretical derivation, and then the effects of different parameters on the photon blockade effect of the system can be studied.
\section{Analytic result}
The wave function of the system is expanded by using the Fork basis state, truncated to at most two photon states, then the wave function $\left| \psi \right\rangle$ of the system can be expressed as
 \begin{equation}\label{eqn7}
\left| \psi \right\rangle = {C_0}\left| 0 \right\rangle + {C_1}\left| 1 \right\rangle + {C_2}\left| 2 \right\rangle ,
\end{equation}
where $\left| \psi \right\rangle$ is the quantum state of the photon, $C_0$, $C_1$ and $C_2$ are the probability amplitudes of the quantum states $\left| 0 \right\rangle$, $\left|1\right\rangle$  and $\left| 2 \right\rangle$, respectively.
Considering the dissipation and attenuation of the system, the non-Hermitian Hamiltonian ${\hat H_{{\text{non}}}}$ of the system can be expressed as

\begin{equation}\label{eqn8}
  \begin{split} 
{\hat H_{{\text{non}}}} =\;& \varDelta {\hat a^\dagger }\hat a + U{\hat a^\dagger }{\hat a^\dagger }\hat a\hat a + {\text{i}}G\left( {{{\hat a}^\dagger }{{\hat a}^\dagger } - \hat a\hat a} \right) \\
&+ F\left( {{{\hat a}^\dagger }{{\text{e}}^{{\text{i}}\phi }} + \hat a{{\text{e}}^{ - {\text{i}}\phi }}} \right) - {\text{i}}\frac{\kappa }{2}{\hat a^\dagger }\hat a.
\end{split}
\end{equation}

Substituting the wave function $\left| \psi \right\rangle$  of the system and the non-Hermitian Hamiltonian ${\hat H_{{\text{non}}}}$ of the system into the Schrödinger equation ${\text{i}}\dfrac{{\partial \left| \psi \right\rangle }}{{\partial t}} = {\hat H_{{\text{non}}}}\left| \psi \right\rangle$, from the equality of the coefficients for the same quantum state, we obtain:
\begin{equation}\label{eqn9}
 \left\{\begin{aligned}
&\text{i}\frac{\partial {C}_{0}}{\partial t}=F{\text{e}}^{-\text{i}\phi}{C}_{1}-\text{i}\sqrt{2}G{C}_{2},\\ 
&\text{i}\frac{\partial {C}_{1}}{\partial t}=F{\text{e}}^{\text{i}\phi }{C}_{0}   +  \left(\varDelta -\text{i}\frac{\kappa }{2}\right){C}_{1}   +  \sqrt{2}F{\text{e}}^{-\text{i}\phi }{C}_{2},\\
&\text{i}\frac{\partial {C}_{2}}{\partial t}  =  \text{i}\sqrt{2}G{C}_{0}   +  \sqrt{2}F{\text{e}}^{\text{i}\phi }{C}_{1}   +  (2\varDelta  - \text{i}\kappa  + 2U ){C}_{2}.
\end{aligned}\right.
\end{equation}

For the steady state, the partial derivative of the probability amplitude of the quantum state with respect to time is zero, and we get:
\begin{equation}\label{eqn10}
\left\{\begin{aligned}
&0=F{\text{e}}^{-\text{i}\phi }{C}_{1}-\text{i}\sqrt{2}G{C}_{2},\\
&0=F{\text{e}}^{\text{i}\phi }{C}_{0}+\left(\varDelta -\text{i}\frac{\kappa }{2}\right){C}_{1}+\sqrt{2}F{\text{e}}^{-\text{i}\phi }{C}_{2},\\ 
&0   =  \text{i}\sqrt{2}G{C}_{0}   +  \sqrt{2}F{\text{e}}^{\text{i}\phi }{C}_{1}   +  (2\varDelta -\text{i}\kappa  +  2U ){C}_{2}.
\end{aligned}\right.
\end{equation}

According to the characteristics of photon distribution, the probability amplitude of its quantum state satisfies $C_0 \gg C_1 \gg C_2$, and considering the condition that the external driving is weak, the first formula in the equations (10) can be regarded as approximately valid. For the convenience of calculation, let $C_0$ be a constant, and assume that $C_0 \approx 1$, then there is
\begin{equation}\label{eqn11}
\left\{\begin{aligned}
&{C}_{1}=\frac{2{C}_{0}F [(2\varDelta -\text{i}\kappa +2U ){\text{e}}^{\text{i}\phi }-2{\mathrm{i}}{\text{e}}^{-\text{i}\phi }G ]}{4{F}^{2} - (2\varDelta -\text{i}\kappa) (2\varDelta -\text{i}\kappa +2U )},\\
&{C}_{2}=-\frac{\sqrt{2}{C}_{0}\left(2{F}^{2}{\text{e}}^{2\text{i}\phi }-G\kappa -2\text{i}\varDelta G\right)}{4{F}^{2}-\left(2\varDelta -\text{i}\kappa \right)\left(2\varDelta -\text{i}\kappa +2U\right)}.\end{aligned} \right.
\end{equation}

For the photon blockade of a single-mode cavity, one has $C_2=0$, i.e.

\begin{equation*}
2{F^2}{{\text{e}}^{2{\text{i}}\phi }} - G\kappa - 2{\text{i}}\varDelta G = 0 .
\end{equation*}

Using Euler's formula ${{\text{e}}^{{\text{i}}\theta }} = \cos \theta + {\text{i}}\sin \theta$, it is rewritten as
\begin{equation*}
2{F^2}\left( {\cos 2\phi + {\text{i}}\sin 2\phi } \right) - G\kappa - 2{\text{i}}\varDelta G = 0 .
\end{equation*}

To make this equation equal to zero, both the real part and the imaginary part are zero, that is,
\begin{equation}\label{eqn12}
\left\{\begin{aligned}
&2{F}^{2}\cos 2\phi -G\kappa =0,\\
&2{F}^{2}\sin 2\phi -2\varDelta G=0.\end{aligned}\right.
\end{equation}

Thus, the optimal condition for photon blockade of the system is
\begin{equation}\label{eqn13}
G = \frac{{2{F^2}\left( {\cos 2\phi + \sin 2\phi } \right)}}{{\kappa + 2\varDelta }}.
\end{equation}

\section{Comparison and discussion between numerical and analytical results}
In order to investigate the photon blockade of the system, the numerical results of the equal-time second-order correlation function ${g^{\left( 2 \right)}}\left( 0 \right)$  in the system are studied by numerical simulation. In the numerical calculation procedure, the effective Hamiltonian derived from Eq. (3) is employed to numerically solve for the steady-state solution of the master equation given by Eq. (4). Based on the obtained steady-state density matrix, the equal-time second-order correlation function is evaluated according to Eq. (5), and the mean photon number is computed via Eq. (6). All numerical simulations presented in this work are performed using the open-source computational package Quantum Optics Toolbox for MATLAB\cite{ref44,ref45}. For convenience, the cavity decay rate $\kappa$ is adopted as the reference unit to normalize all other physical quantities.  
\subsection{Comparison between numerical results of equal-time second-order correlation function and analytical results of photon-blockade optimum condition}
\begin{figure*}
\includegraphics[width=1.00\textwidth]{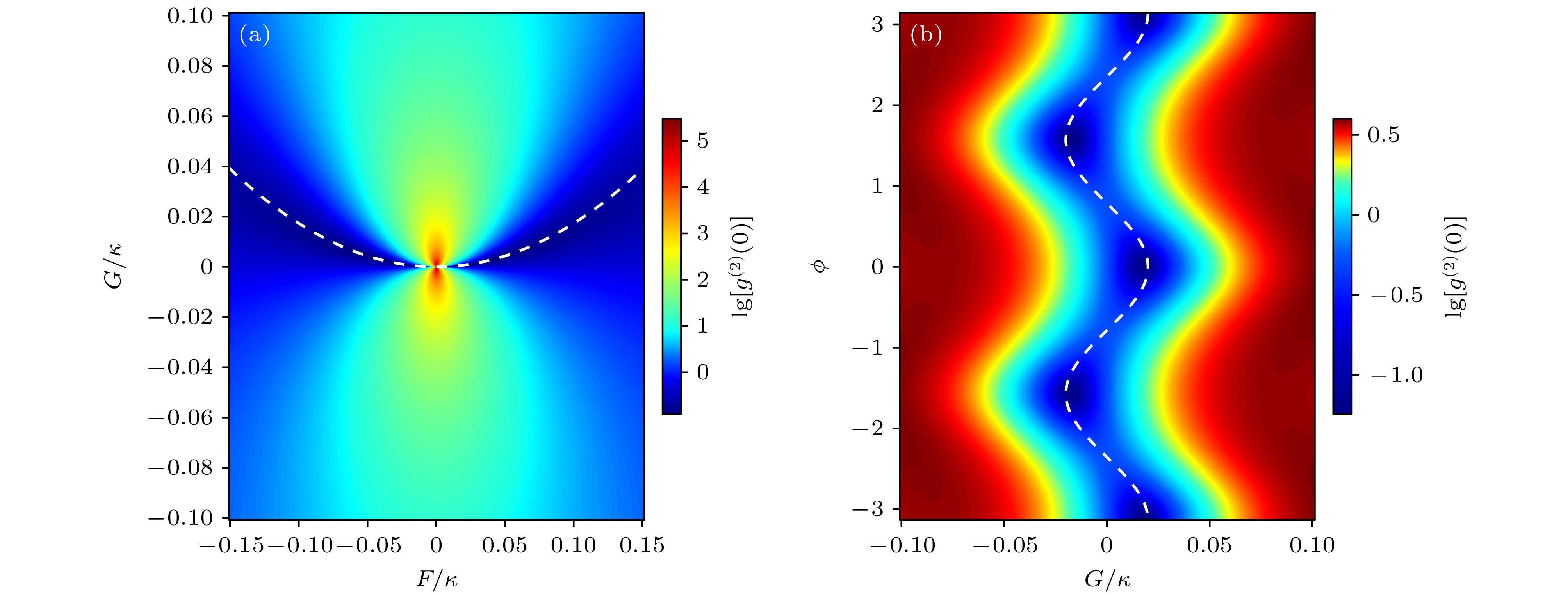}%
\caption{\label{fig:fig1} Logarithmic value of the equal-time second-order correlation function ${g^{\left( 2 \right)}}\left( 0 \right)$ versus different physical parameters are presented: (a) Logarithmic value of ${g^{\left( 2 \right)}}\left( 0 \right)$ as a function of the driving strength $F/\kappa$ and the optical parametric amplifier nonlinear coefficient $G/\kappa$ of the optical parametric amplifier, where$\phi = {\pi} /12$ and $U/\kappa=0.5$; (b) logarithmic value of ${g^{\left( 2 \right)}}\left( 0 \right)$ as a function of the optical parametric amplifier nonlinear coefficient $G/\kappa$ and the driving phase $\phi$, where $F/\kappa = 0.1$ and $U/\kappa=0.5$. In both figures, the white dashed lines, derived from Eq. (13), indicate the analytical solutions corresponding to the optimal conditions for photon blockade.}
\end{figure*}
The numerical results of the equal-time second-order correlation function and the analytical results of the optimum condition of photon blockade are shown in Fig.~\ref{fig:fig1}. The logarithmic value of the isochronous second-order correlation function ${g^{\left( 2 \right)}}\left( 0 \right)$ is given Fig.~\ref{fig:fig1}(a) as a function of the driving strength $F/\kappa$ and the nonlinear coefficient of the optical parametric amplifier $G/\kappa$. The other parameters used in the numerical calculation are set to $\phi = {\pi} /12$ and $U/\kappa=0.5$. By analyzing Fig.~\ref{fig:fig1}(a), it can be observed that there is a region of ${g^{\left( 2 \right)}}\left( 0 \right)<1$ (i.e., $\lg \left[ {{g^{(2)}}(0)} \right] < 0$) in a specific parameter interval, which clearly indicates that the system is in a photon blockade state\cite{ref46}. The white dashed line in Fig.~\ref{fig:fig1}(a) is drawn based on equation (12), which represents the theoretical analytical solution of the optimal condition of photon blockade. A comparison reveals that, the white dashed line is consistent with the distribution of the darkest region of the corresponding color mapping in the numerical calculation, which indicates that the theoretical analytical solution of the optimal condition of photon blockade is highly consistent with the minimum distribution in the numerical calculation. In Fig.~\ref{fig:fig1}(b), it shows the variation of the logarithmic value of the equal-time second-order correlation function ${g^{\left( 2 \right)}}\left( 0 \right)$ with respect to the optical parametric amplifier nonlinear coefficient $G/\kappa$ and the driving force phase $\phi $. The other parameters used in the numerical calculation are set to $F/\kappa=0.1$ and $U/\kappa=0.5$. A prominent left-right oscillating dark band region can be observed from the Fig.~\ref{fig:fig1}(b), which corresponds to the region where the equal-time second-order correlation function satisfies the ${g^{\left( 2 \right)}}\left( 0 \right)<1$. This non-classical feature clearly indicates that the system is in the photon blockade state. The white dashed line here is also given by equation (13), and it is also seen that the white dashed line is consistent with the distribution of the dark region, which shows that the theoretical value of the optimal condition of photon blockade is in good agreement with the numerical result.

\subsection{Numerical results of the mean photon number in a photon-blockade system}
As the core parameter to characterize the performance of a single photon source, the average photon number directly determines the effective single photon flux that the source can provide per unit time, and making it a key metric to evaluate the practicability of a single photon source. To investigate the physical factors influencing the brightness of the single-photon source in this system, we employed numerical simulations to systematically study the regulation of the average photon number by varying physical parameters. Specifically, the synergetic effects of the system detuning $\varDelta$, the driving force strength $F$ and phase $\phi$, the optical parametric amplifier nonlinear coefficient $G$, and the Kerr nonlinearity strength $U$ on the average photon number are analyzed.

\begin{figure*}
\includegraphics[width=1.00\textwidth]{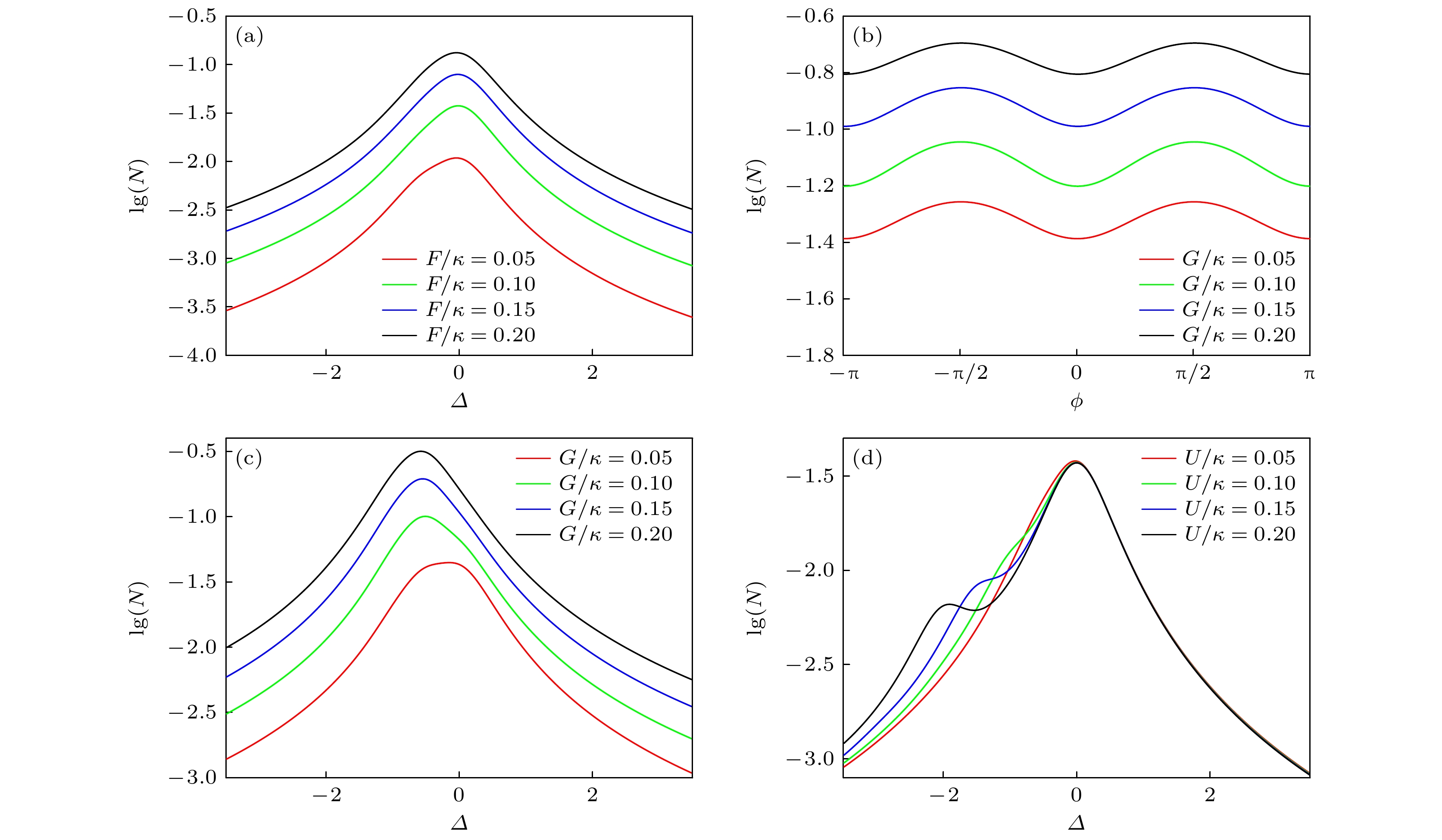}%
\caption{\label{fig:fig2} Logarithmic value of the average photon number $N$ versus different parameters: (a) $\lg(N)$ as a function of detuning $\varDelta/\kappa$ at different driving strengths $F/\kappa$; (b) phase dependence of $\lg(N)$ under varying of the optical parametric amplifier nonlinear coefficients $G/\kappa$; (c) detuning dependence of $\lg(N)$ for different optical parametric amplifier nonlinear coefficients $G/\kappa$; (d) $\lg(N)$ versus detuning $\varDelta/\kappa$ at distinct Kerr nonlinearity strengths $U/\kappa$.}
\end{figure*}

The Fig.~\ref{fig:fig2} shows the logarithmic of the average photon number of the system $\lg(N)$ as a function of different parameters. The Fig.~\ref{fig:fig2}(a) shows the variation of the $\lg(N)$ with the detuning $\varDelta/\kappa$ under different driving strength $F/\kappa$. The results show that for all driving strengths, the average photon number exhibits a significant peak at the detuning of zero, that is, $\varDelta/\kappa = 0$, and the peak intensity increases monotonically with the increase of the driving strength $F/\kappa$. This result reveals that the system can achieve the optimal photon generation efficiency when the driving force frequency matches the eigenfrequency of the system, that is, the system is in resonance. Fig. 2(b) shows the variation of $\lg(N)$ with the phase $\phi$ of the driving force under different optical parametric amplifier nonlinear coefficient $G/\kappa$. It can be found that the average photon number is periodically modulated with the driving force phase, with a maximum at $\phi = \pm {{\pi} \mathord{\left/ {\vphantom {{\pi} 2}} \right. } 2}$ and a modulation period of $\pi$. The peak intensity of the mean photon number is positively correlated with the optical parametric amplifier nonlinear coefficient $G/\kappa$, and the modulation amplitude increases with the increase of the $G/\kappa$. When the optical parametric amplifier nonlinear coefficient $G/\kappa$ is fixed, the fluctuation range of the average photon number of the system is relatively limited, which indicates that there is a certain saturation effect on the regulation of the photon number distribution by the driving force phase. In Fig.~\ref{fig:fig2}(c), the relationship between $\lg(N)$ and detuning $\varDelta/\kappa$ is studied under different optical parametric amplifier nonlinear coefficient $G/\kappa$. It can be concluded from Fig.~\ref{fig:fig2}(c) that the average photon number curve exhibits a single-peak characteristic, and the peak position shows an obvious nonlinear dependence: when the optical parametric amplifier nonlinear coefficient is small, such as $G/\kappa$=0.05, the peak position corresponds to the detuning $\varDelta/\kappa$=0, while with the increase of the optical parametric amplifier nonlinear coefficient, the peak position shifts to the negative detuning direction. The influence of Kerr nonlinearity strength $U/\kappa$ on the average photon number of the system is studied, as shown in Fig.~\ref{fig:fig2}(d). The results show that the peak value of the average photon number is always stable at $\varDelta/\kappa$=0. It can also be found that the four curves are almost coincident when the Kerr nonlinearity strength $U/\kappa$ is different. This result shows that the influence of Kerr nonlinearity on the average photon number of the system is relatively small, and also confirms that the average photon number of the system has robust stability when the system is on resonant conditions. The above numerical results can provide a theoretical reference for the performance optimization of single photon source, and the analysis of related parameters may be of reference value for the selection of experimental conditions.
\subsection{Effect of driving force phase on photon blockade in the system}
\begin{figure*}
\includegraphics[width=1.00\textwidth]{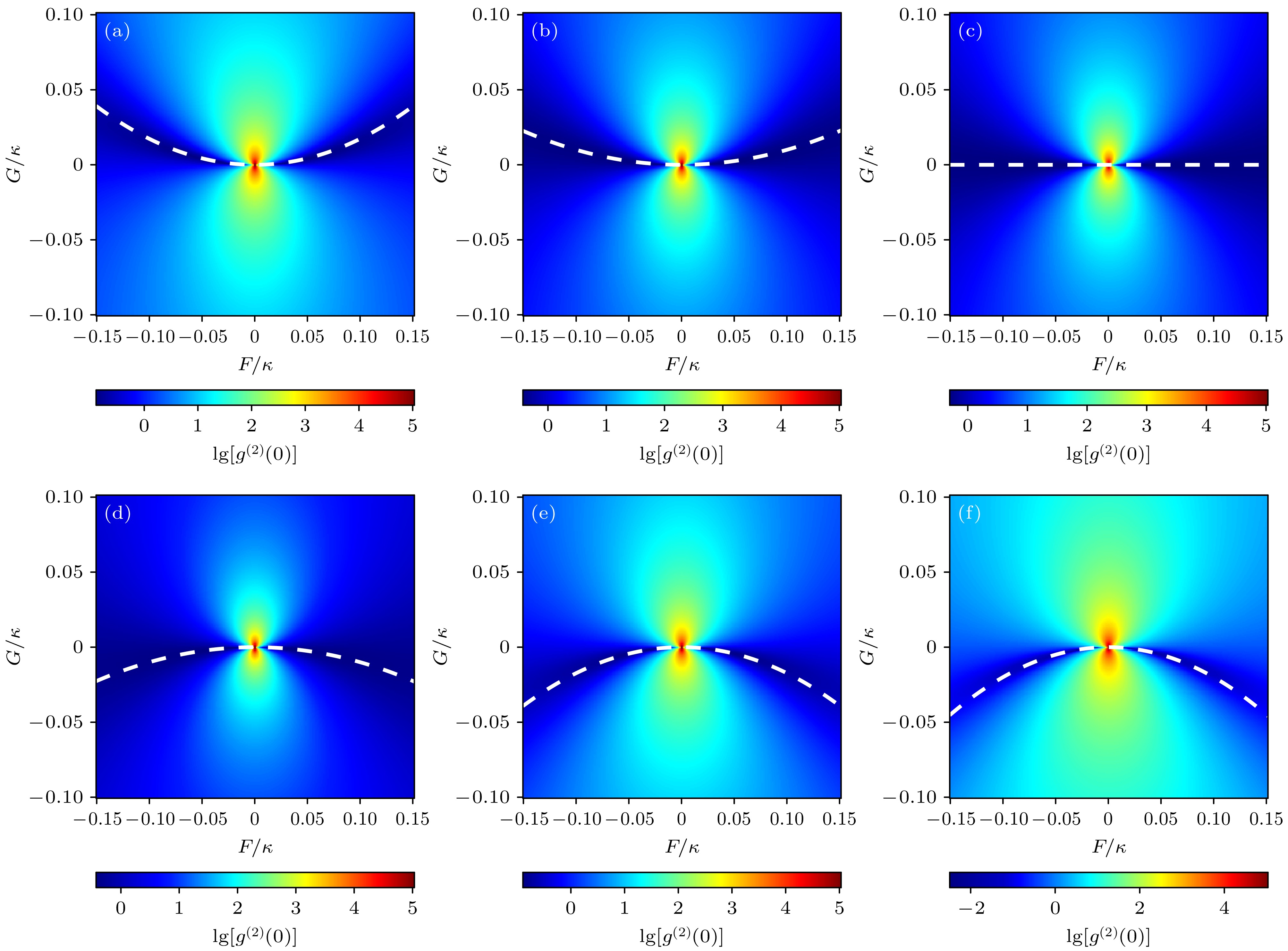}%
\caption{\label{fig:fig3} Logarithmic value of ${g^{\left( 2 \right)}}\left( 0 \right)$ as a function of the driving strength $F/\kappa$ and the nonlinear coefficient $G/\kappa$ of the optical parametric amplifier under different driving phases $\phi$: (a) $\phi =\pi/12$; (b) $\phi =\pi/6$; (c) $\phi =\pi/4$; (d) $\phi =\pi/3$; (e) $\phi =5\pi/12$; (f) $\phi =\pi/2$. In all panels, the white dashed lines, derived from Eq. (13), represent the analytical solutions for the optimal photon blockade conditions. The Kerr nonlinearity strength was consistently set to $U/\kappa=0.5$ in the numerical simulations.}
\end{figure*}

In this section, we study the modulation mechanism of the photon blockade effect by adjusting the driving force phase $\phi$ under the resonance condition, that is, $\varDelta/\kappa$=0. The Fig.~\ref{fig:fig3} shows the contour plots of the logarithmic value of the equal-time second-order correlation function ${g^{\left( 2 \right)}}\left( 0 \right)$ as a function of the driving force strength $F/\kappa$ and the optical parametric amplifier nonlinear coefficient $G/\kappa$, where different subplots correspond to different driving force phase settings, and in all plots, the Kerr nonlinearity strength is set to $U/\kappa$=0.5. It can be seen from the Fig.~\ref{fig:fig3} that photon blockade can exist in the system in each case, and the theoretical value of the optimal condition of photon blockade is in good agreement with the numerical results.

It is worth noting that the change of the driving force phase will significantly affect the range of parameter space where the photon blockade effect appears. The contour plots of the logarithm of the equal-time second-order correlation function ${g^{\left( 2 \right)}}\left( 0 \right)$ as a function of the driving force strength $F/\kappa$ and the optical parametric amplifier nonlinear coefficient $G/\kappa$ are shown in the Fig.~\ref{fig:fig3}(a) —(f), with different driving force phase $\phi =\pi/12$, $\pi/6$,  $\pi/4$,  $\pi/3$, $5\pi/12$ and $\pi/2$, respectively. It can be seen from Fig.~\ref{fig:fig3} that the phase of the driving force changes from $\phi =\pi/12$ to $\phi =\pi/4$, and the corresponding subfigures is from Fig.~\ref{fig:fig3}(a) to(c). The optimal photon blockade region shifts significantly in the $F-G$ parameter plane, and evolves from a parabolic region with an upward opening to a horizontally symmetrical zonal distribution. As the phase of the driving force increases to $\phi =\pi/3$, as shown in Fig.~\ref{fig:fig3}(d), the direction of the parabolic opening reverses, and the optimal photon blocking region changes to a parabolic region with a downward opening. Furthermore, when the driving force phase is increased to $\phi =5\pi/12$ and  $\pi/2$, the optimal photon blocking region still opens downward, and the difference between them is that the range of the photon blocking region is reduced, as shown in Fig.~\ref{fig:fig3}(e) and Fig.~\ref{fig:fig3}(f). The optimal condition for photon blockade given by Eq. (13) is given by the white dashed line in the figure. It can be observed that the theoretical values of the optimal condition agree well with the results obtained from numerical simulations. These results, from both numerical simulation and theoretical analysis, confirm the control effect of the driving force phase on the photon blockade effect. This driving force phase-dependent photon blockade control mechanism provides a new manipulation dimension for programmable quantum devices, such as dynamic single photon sources, which can significantly improve the flexibility and integration of quantum information processing by realizing the topological reconstruction of parameter space through phase regulation\cite{ref24,ref47,ref48}.

\subsection{Effect of Kerr nonlinearity on photon blockade in the system}
\begin{figure*}
\includegraphics[width=1.00\textwidth]{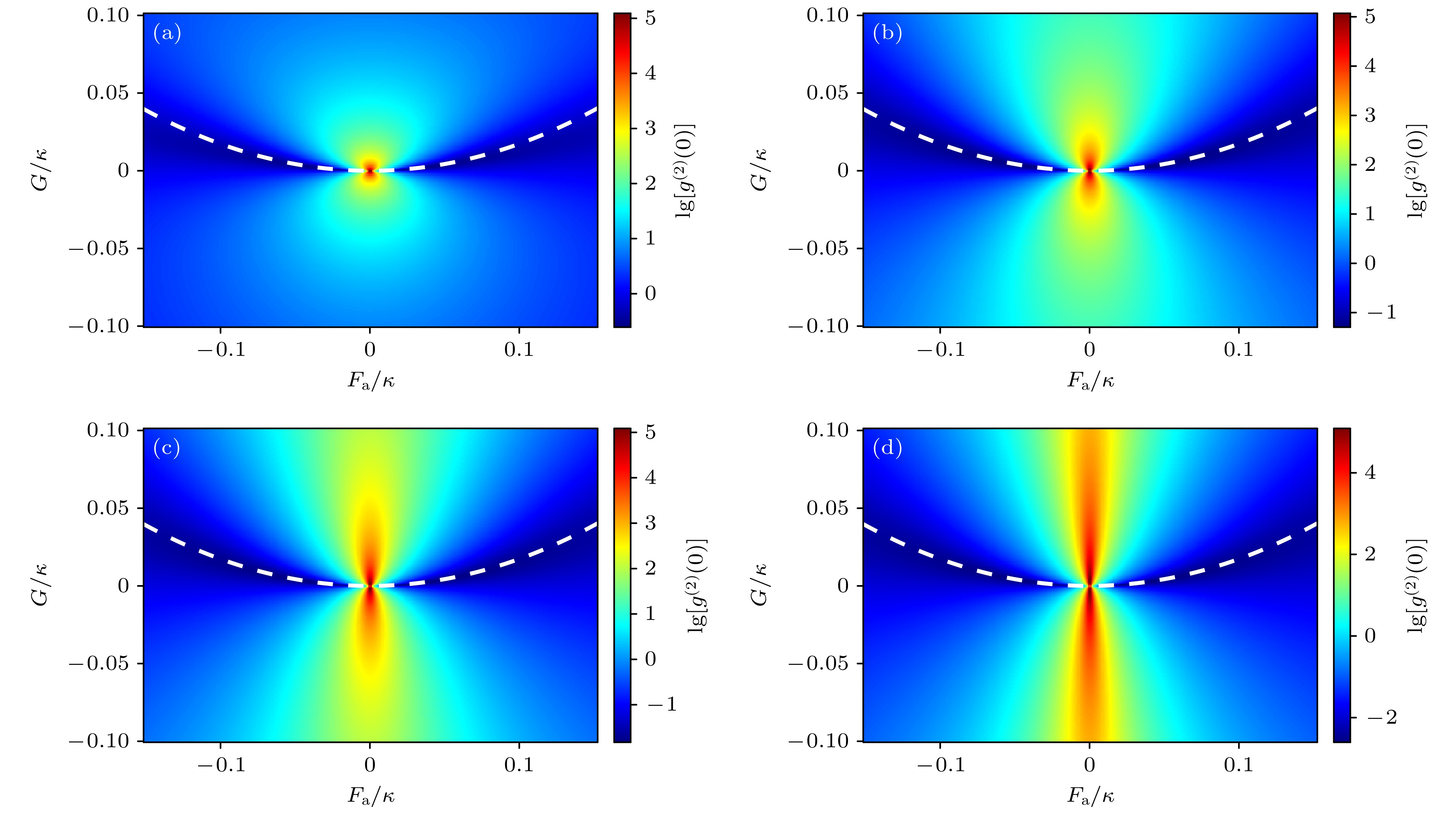}%
\caption{\label{fig:fig4} Logarithmic value of ${g^{\left( 2 \right)}}\left( 0 \right)$ as a function of the driving strength $F/\kappa$ and the nonlinear coefficient $G/\kappa$ of the optical parametric amplifier under different Kerr nonlinearity strength $U/\kappa$: (a) $U/\kappa$=0.1; (b) $U/\kappa$=1.0; (c) $U/\kappa$=2.0; (d) $U/\kappa$=5.0. In all panels, the white dashed lines, derived from Eq. (13), represent the analytical solutions for the optimal photon blockade conditions.}
\end{figure*}

In order to investigate the influence of Kerr nonlinearity on the photon blockade in the system, the distribution characteristics of the equal-time second-order correlation function with different Kerr nonlinearity strengths are calculated by numerical simulation. The Fig.~\ref{fig:fig4}(a) —(d) shows the distribution of the logarithmic value of the equal-time second-order correlation function $\lg{g^{\left( 2 \right)}}\left( 0 \right)$ in the parameter space composed of the driving force strength $F/\kappa$ and the optical parametric amplifier nonlinear coefficient $G/\kappa$ when the Kerr nonlinearity strength $U/\kappa$ set at 0.1, 1.0, 2.0 and 5.0, respectively, and the color depth of the contour line represents the value of the $\lg{g^{\left( 2 \right)}}\left( 0 \right)$. By comparing the numerical simulation results from Fig.~\ref{fig:fig4}(a) to (d), it can be found that there is a significant photon blockade effect in the system in a wide parameter range of the Kerr nonlinearity strength  $U/\kappa$ from 0.1 to 5.0. It is worth noting that the position and shape of the optimal photon blockade region show good stability. This numerical result is in good agreement with the analytical result of the optimal photon blockade conditions in Eq. (13).

It is worth mentioning that the universal photon blockade proposed by Zhou \emph{et al.}\cite{ref49} recently breaks through the classification limitation of the traditional photon blockade and realizes the photon blockade effect independent of the nonlinear strength. Through the above theoretical analysis and numerical simulation, it can be found that in the system designed in this study, the Kerr nonlinearity has little effect on the optimal photon blockade region of the system. This characteristic enables the system to achieve efficient photon blockade in a wide range of nonlinear parameters, thus showing a typical universal photon blockade feature.

\section{Physical mechanism of photon blockade in the system}
\begin{figure}
\includegraphics[width=0.5\textwidth]{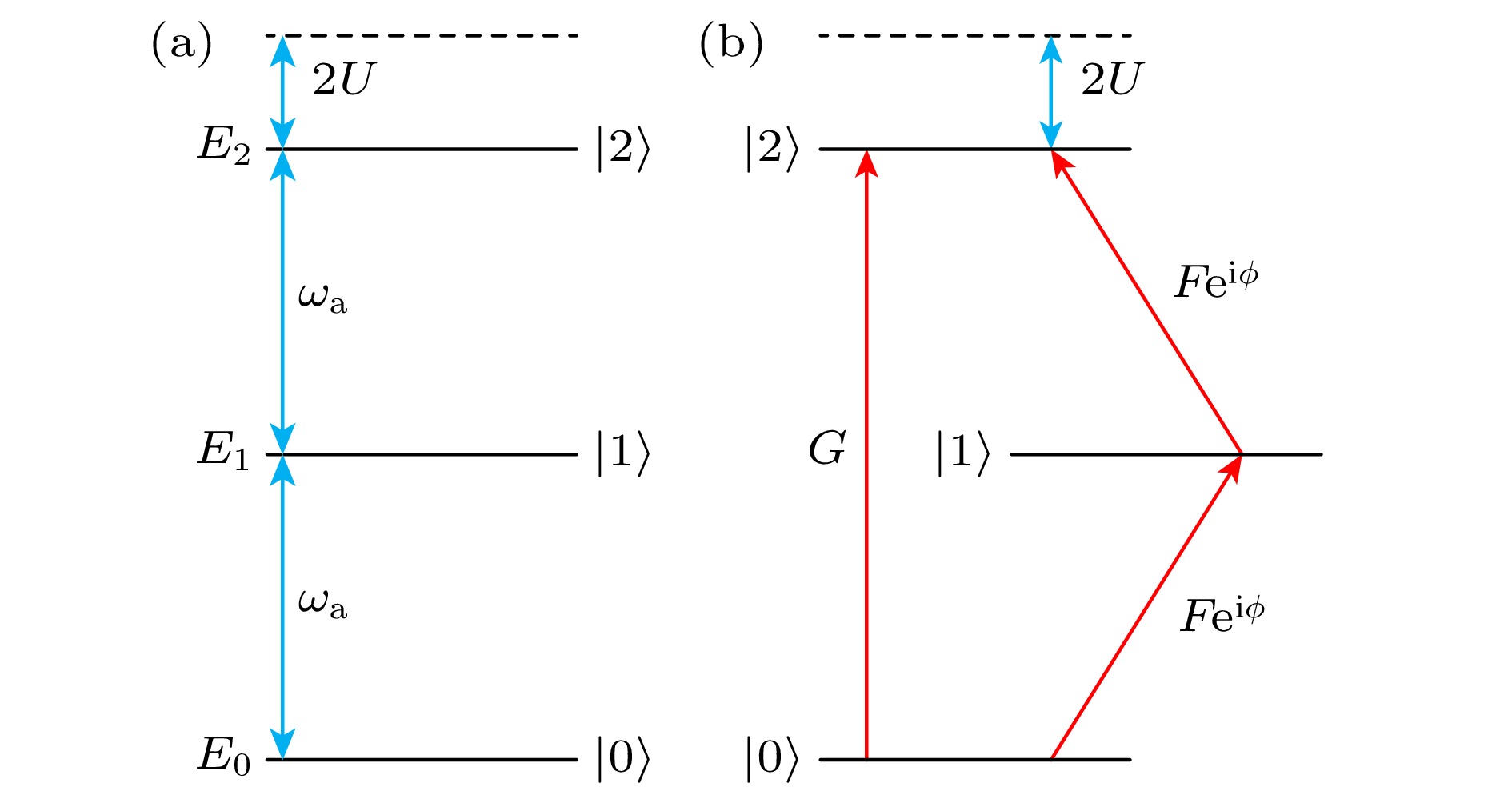}%
\caption{\label{fig:fig5} Schematic diagram of the system energy-level and the transition paths between different photon states: (a) Energy level diagram; (b) photon state transition pathways.}
\end{figure}

In this section, the physical mechanism of photon blockade in the system consisting of a single mode cavity and an optical parametric amplifier is discussed. When the external driving is not considered, the intrinsic Hamiltonian of the system is ${\hat H_0} = {\omega _a}{a^\dagger }a + U{a^\dagger }{a^\dagger }aa$, and the energy level of the system can be expressed as ${E_n} = n{\omega _a} + n\left( {n - 1} \right)U$. The energy levels of the system and the transition paths of photons between different energy levels are given Fig.~\ref{fig:fig5}. The Fig.~\ref{fig:fig5}(a) shows the energy level diagram of the system. The Fig.~\ref{fig:fig5}(b) shows the possible transition paths of the photon state of the system. From the Fig.~\ref{fig:fig5}(b), it can be found that there are two paths for the system to reach the two-photon state. 1) Direct path: under the action of the optical parametric amplifier, the ground state is directly transitioned to the two-photon state, and the path is $\;\left| 0 \right\rangle \xrightarrow{G}\;\left| 2 \right\rangle$; 2) Indirect path: under the action of external driving force, it first transitions from the ground state to the single-photon state, and then continues to transition from the single-photon state to the two-photon state under the action of external driving force $\;\left| 0 \right\rangle \xrightarrow{{F{{\text{e}}^{{\text{i}}\phi }}}} \left| 1 \right\rangle \;\xrightarrow{{F{{\text{e}}^{{\text{i}}\phi }}}}\left| 2 \right\rangle$. When the parameters in the system satisfy the optimal photon blockade condition given by Eq. (13), the photons from two different transition paths will produce a completely destructive interference effect, which significantly reduces the probability of two-photon excitation, thus showing a strong photon antibunching characteristic in the experiment and achieving efficient photon blockade. Although the Kerr nonlinearity can significantly change the energy level structure of the system and lead to the nonlinear broadening of the energy level spacing, it does not directly affect the interference characteristics of the photon transition path. This important feature enables the system to maintain a stable photon blockade effect over a wide range of Kerr nonlinearity strength.

In summary, the physical mechanism underlying the photon blockade effect in the system arises from the synergistic interaction between the nonlinear effects of the optical parametric amplifier and the external driving field through quantum interference. This cooperation effectively suppresses the population of the two-photon excitation state, thereby achieving significant photon antibunching, i.e., photon blockade. The essence of this photon blockade phenomenon originates from the selective suppression of multiphoton transitions by destructive interference between quantum paths under the condition of system parameter optimization.

\section{Conclusion} 
By combining analytical solution with numerical simulation, the control mechanism of photon blockade effect in a hybrid quantum system composed of a Kerr-medium single-mode cavity and an optical parametric amplifier (OPA) is studied, which is influenced by the nonlinear coefficient of OPA, the driving force strength, the driving force phase and the Kerr nonlinearity strength.

In order to study the photon blockade of the system, a master equation describing the dynamic process of the system is established based on the effective Hamiltonian of the system with the consideration of the decay of the single mode cavity. In order to obtain the analytical solution of the optimal condition of photon blockade, the quantum state of the system is expanded into a two-photon state by using the Fork ground state, and the analytical expression of the probability amplitude of the quantum state of the system in the steady state is derived by solving the Schrödinger equation of the system, and then the analytical result of the optimal condition of photon blockade is obtained. The numerical results of the equal-time second-order correlation function in the system are calculated by numerical simulation and compared with the analytical results of the optimal condition of photon blockade. The results show that the photon blockade can exist in the system under the condition of appropriate parameters, and the analytical results of the optimal condition of photon blockade are highly consistent with the numerical results, which not only verifies the correctness of the analytical results, but also confirms the effectiveness of the photon blockade condition in the system. The numerical simulation results show that the average photon number can be observed to increase significantly under the resonance condition, which provides a potential theoretical support for optimizing the brightness parameters of single photon sources. Then, the modulation effect of the phase of the driving force on the photon blockade in the system is studied, and it is found that the change of the phase of the driving force will significantly affect the parameter space range where the photon blockade effect occurs, and the optimal photon blockade region will be significantly shifted in the plane of the nonlinear coefficient $F$ and the optical parametric amplifier nonlinear coefficient $G$. Furthermore, the orientation of the parabolic opening in the optimal photon blockade region undergoes a reversal: starting as an upward-opening parabola, transitioning through a horizontal band-like intermediate state, and ultimately evolving into a downward-opening parabolic structure. These results are consistent with the analytical results of the optimal condition of photon blockade, and both the numerical and theoretical results confirm the control effect of the driving force phase on the photon blockade effect. The influence of Kerr nonlinearity on the photon blockade in the system is further discussed, and the results show that there is a significant photon blockade effect in the system in a wide parameter range of the Kerr nonlinearity, showing a typical universal photon blockade feature.

Finally, the physical mechanism of the photon blockade phenomenon in the system is discussed. According to the energy level of the system and the related interaction, it is known that there are two photon transition paths to reach the two-photon state in the system. Under the condition of appropriate parameters, the photons of the two different transition paths in the system produce completely destructive quantum interference effect, which leads to the probability of the two-photon state close to zero and effectively suppresses the two-photon excitation, that is, the photon blockade phenomenon occurs. The existence of Kerr nonlinearity only changes the energy level structure of the system, but does not directly affect the photon transition path, which makes the system have photon blockade phenomenon in a wide range of Kerr nonlinearity strength.

\begin{acknowledgments}
This work is supported by the Science and Technology Research Project of Henan Province, China (Grant No. 242102231052).

I would like to thank Dr. Zhou Yanhui of the Quantum Information Research Center of Shangrao Normal University for his helpful discussions and careful guidance.
\end{acknowledgments}

\bibliography{mybib}

@article{ref1,
author = {Zubizarreta Casalengua, Eduardo and López Carreño, Juan Camilo and Laussy, Fabrice P. and Valle, Elena del},
title = {Conventional and Unconventional Photon Statistics},
journal = {Laser \& Photonics Reviews},
volume = {14},
number = {6},
pages = {1900279},
doi = {https://doi.org/10.1002/lpor.201900279},
year = {2020}
}

@article{ref2,
  title = {Chiral Interaction Induced Near-Perfect Photon Blockade},
  author = {Lu, Zhi-Guang and Wu, Ying and L\"u, Xin-You},
  journal = {Phys. Rev. Lett.},
  volume = {134},
  issue = {1},
  pages = {013602},
  numpages = {7},
  year = {2025},
  month = {Jan},
  publisher = {American Physical Society},
  doi = {10.1103/PhysRevLett.134.013602},
  url = {https://link.aps.org/doi/10.1103/PhysRevLett.134.013602}
}

@article{ref3,
  title={Photon blockade in an optical cavity with one trapped atom},
  author={Birnbaum, Kevin M and Boca, Andreea and Miller, Russell and Boozer, Allen D and Northup, Tracy E and Kimble, H Jeff},
  journal={Nature},
  volume={436},
  number={7047},
  pages={87--90},
  year={2005},
  publisher={Nature Publishing Group UK London},
  doi = {https://doi.org/10.1038/nature03804},
  url = {https://doi.org/10.1038/nature03804}
}

@article{ref4,
  title = {Observation of the Unconventional Photon Blockade},
  author = {Snijders, H. J. and Frey, J. A. and Norman, J. and Flayac, H. and Savona, V. and Gossard, A. C. and Bowers, J. E. and van Exter, M. P. and Bouwmeester, D. and L\"offler, W.},
  journal = {Phys. Rev. Lett.},
  volume = {121},
  issue = {4},
  pages = {043601},
  numpages = {5},
  year = {2018},
  month = {Jul},
  publisher = {American Physical Society},
  doi = {10.1103/PhysRevLett.121.043601},
  url = {https://link.aps.org/doi/10.1103/PhysRevLett.121.043601}
}

@article{ref5,
  title = {Observation of the Unconventional Photon Blockade in the Microwave Domain},
  author = {Vaneph, Cyril and Morvan, Alexis and Aiello, Gianluca and F\'echant, Mathieu and Aprili, Marco and Gabelli, Julien and Est\`eve, J\'er\^ome},
  journal = {Phys. Rev. Lett.},
  volume = {121},
  issue = {4},
  pages = {043602},
  numpages = {5},
  year = {2018},
  month = {Jul},
  publisher = {American Physical Society},
  doi = {10.1103/PhysRevLett.121.043602},
  url = {https://link.aps.org/doi/10.1103/PhysRevLett.121.043602}
}

@article{ref6,
  title={High-efficiency single-photon source above the loss-tolerant threshold for efficient linear optical quantum computing},
  author={Ding, Xing and Guo, Yong-Peng and Xu, Mo-Chi and Liu, Run-Ze and Zou, Geng-Yan and Zhao, Jun-Yi and Ge, Zhen-Xuan and Zhang, Qi-Hang and Liu, Hua-Liang and Wang, Lin-Jun and others},
  journal={Nature Photonics},
  volume = {19},
  pages={387--391},
  year={2025},
  publisher={Nature Publishing Group UK London},
  doi = {10.1038/s41566-025-01639-8},
  url = {https://doi.org/10.1038/s41566-025-01639-8}
}

@article{ref7,
  title = {Conventional photon blockade with a three-wave mixing},
  author = {Zhou, Y. H. and Zhang, X. Y. and Wu, Q. C. and Ye, B. L. and Zhang, Zhi-Qiang and Zou, D. D. and Shen, H. Z. and Yang, Chui-Ping},
  journal = {Phys. Rev. A},
  volume = {102},
  issue = {3},
  pages = {033713},
  numpages = {6},
  year = {2020},
  month = {Sep},
  publisher = {American Physical Society},
  doi = {10.1103/PhysRevA.102.033713},
  url = {https://link.aps.org/doi/10.1103/PhysRevA.102.033713}
}

@article{ref8,
doi = {10.1088/1402-4896/aca2fc},
url = {https://doi.org/10.1088/1402-4896/aca2fc},
year = {2023},
month = {feb},
publisher = {IOP Publishing},
volume = {98},
number = {3},
pages = {035108},
author = {Wang, Zhu-Xin and Yang, Hui and Wang, Xiao-Qian and Lin, Hong-Yu and Yao, Zhi-Hai},
title = {Conventional photon blockade in a four-wave mixing system with Kerr nonlinearity},
journal = {Physica Scripta}
}

@article{ref9,
author = {Hongyu Lin and Xiaoqian Wang and Zhihai Yao and Dandan Zou},
journal = {Opt. Express},
number = {12},
pages = {17643--17652},
publisher = {Optica Publishing Group},
title = {Kerr-nonlinearity enhanced conventional photon blockade in a second-order nonlinear system},
volume = {28},
month = {Jun},
year = {2020},
url = {https://opg.optica.org/oe/abstract.cfm?URI=oe-28-12-17643},
doi = {10.1364/OE.385981},
}

@article{ref10,
  title = {Origin of strong photon antibunching in weakly nonlinear photonic molecules},
  author = {Bamba, Motoaki and Imamo\ifmmode \breve{g}\else \u{g}\fi{}lu, Atac and Carusotto, Iacopo and Ciuti, Cristiano},
  journal = {Phys. Rev. A},
  volume = {83},
  issue = {2},
  pages = {021802},
  numpages = {4},
  year = {2011},
  month = {Feb},
  publisher = {American Physical Society},
  doi = {10.1103/PhysRevA.83.021802},
  url = {https://link.aps.org/doi/10.1103/PhysRevA.83.021802}
}

@article{ref11,
  title = {Unconventional photon blockade},
  author = {Flayac, H. and Savona, V.},
  journal = {Phys. Rev. A},
  volume = {96},
  issue = {5},
  pages = {053810},
  numpages = {13},
  year = {2017},
  month = {Nov},
  publisher = {American Physical Society},
  doi = {10.1103/PhysRevA.96.053810},
  url = {https://link.aps.org/doi/10.1103/PhysRevA.96.053810}
}

@article{ref12,
  title = {Unconventional photon blockade with non-Markovian effects in driven dissipative coupled cavities},
  author = {Shen, H. Z. and Yang, J. F. and Yi, X. X.},
  journal = {Phys. Rev. A},
  volume = {109},
  issue = {4},
  pages = {043714},
  numpages = {23},
  year = {2024},
  month = {Apr},
  publisher = {American Physical Society},
  doi = {10.1103/PhysRevA.109.043714},
  url = {https://link.aps.org/doi/10.1103/PhysRevA.109.043714}
}

@article{ref13,
  title = {Photon blockade in non-Hermitian optomechanical systems with nonreciprocal couplings},
  author = {Sun, J. Y. and Shen, H. Z.},
  journal = {Phys. Rev. A},
  volume = {107},
  issue = {4},
  pages = {043715},
  numpages = {16},
  year = {2023},
  month = {Apr},
  publisher = {American Physical Society},
  doi = {10.1103/PhysRevA.107.043715},
  url = {https://link.aps.org/doi/10.1103/PhysRevA.107.043715}
}

@article{ref14,
  title = {Single Photons from Coupled Quantum Modes},
  author = {Liew, T. C. H. and Savona, V.},
  journal = {Phys. Rev. Lett.},
  volume = {104},
  issue = {18},
  pages = {183601},
  numpages = {4},
  year = {2010},
  month = {May},
  publisher = {American Physical Society},
  doi = {10.1103/PhysRevLett.104.183601},
  url = {https://link.aps.org/doi/10.1103/PhysRevLett.104.183601}
}

@article{ref15,
  title = {Strongly Interacting Photons in a Nonlinear Cavity},
  author = {Imamo\ifmmode \bar{g}\else \={g}\fi{}lu, A. and Schmidt, H. and Woods, G. and Deutsch, M.},
  journal = {Phys. Rev. Lett.},
  volume = {79},
  issue = {8},
  pages = {1467--1470},
  numpages = {0},
  year = {1997},
  month = {Aug},
  publisher = {American Physical Society},
  doi = {10.1103/PhysRevLett.79.1467},
  url = {https://link.aps.org/doi/10.1103/PhysRevLett.79.1467}
}

@article{ref16,
  title = {Tunable unconventional phonon blockade in Fabry-Perot cavity and optical parametric amplifier composite system},
  author = {Li Hong and Zhang Si-Qi and Guo Ming and Li Mei-Xuan and Song Li-Jun},
  journal = {Acta Phys. Sin.},
  volume = {68},
  issue = {12},
  pages = {124203},
  numpages = {0},
  year = {2019},
  doi = {10.7498/aps.68.20190154},
  url = {https://doi.org/10.7498/aps.68.20190154}
}

@article{ref17,
  title = {Single-Mode Photon Blockade Enhanced by Bi-Tone Drive},
  author = {Li, Ming and Zhang, Yan-Lei and Wu, Shu-Hao and Dong, Chun-Hua and Zou, Xu-Bo and Guo, Guang-Can and Zou, Chang-Ling},
  journal = {Phys. Rev. Lett.},
  volume = {129},
  issue = {4},
  pages = {043601},
  numpages = {7},
  year = {2022},
  month = {Jul},
  publisher = {American Physical Society},
  doi = {10.1103/PhysRevLett.129.043601},
  url = {https://link.aps.org/doi/10.1103/PhysRevLett.129.043601}
}

@article{ref18,
  title = {Photon Blockade in the Ultrastrong Coupling Regime},
  author = {Ridolfo, A. and Leib, M. and Savasta, S. and Hartmann, M. J.},
  journal = {Phys. Rev. Lett.},
  volume = {109},
  issue = {19},
  pages = {193602},
  numpages = {5},
  year = {2012},
  month = {Nov},
  publisher = {American Physical Society},
  doi = {10.1103/PhysRevLett.109.193602},
  url = {https://link.aps.org/doi/10.1103/PhysRevLett.109.193602}
}

@article{ref19,
  title = {Zero eigenvalues of a photon blockade induced by a non-Hermitian Hamiltonian with a gain cavity},
  author = {Zhou, Y. H. and Shen, H. Z. and Zhang, X. Y. and Yi, X. X.},
  journal = {Phys. Rev. A},
  volume = {97},
  issue = {4},
  pages = {043819},
  numpages = {6},
  year = {2018},
  month = {Apr},
  publisher = {American Physical Society},
  doi = {10.1103/PhysRevA.97.043819},
  url = {https://link.aps.org/doi/10.1103/PhysRevA.97.043819}
}

@article{ref20,
author = {Hongyan Zhu and Xiaomiao Li and Zigeng Li and Fan Wang and Xiaolan Zhong},
journal = {Opt. Express},
number = {13},
pages = {22030--22039},
publisher = {Optica Publishing Group},
title = {Strong antibunching effect under the combination of conventional and unconventional photon blockade},
volume = {31},
month = {Jun},
year = {2023},
url = {https://opg.optica.org/oe/abstract.cfm?URI=oe-31-13-22030},
doi = {10.1364/OE.493612},
}

@article{ref21,
  title = {Tunable photon blockade in coupled semiconductor cavities},
  author = {Shen, H. Z. and Zhou, Y. H. and Yi, X. X.},
  journal = {Phys. Rev. A},
  volume = {91},
  issue = {6},
  pages = {063808},
  numpages = {8},
  year = {2015},
  month = {Jun},
  publisher = {American Physical Society},
  doi = {10.1103/PhysRevA.91.063808},
  url = {https://link.aps.org/doi/10.1103/PhysRevA.91.063808}
}

@article{ref22,
  title = {Unconventional photon blockade with second-order nonlinearity},
  author = {Zhou, Y. H. and Shen, H. Z. and Yi, X. X.},
  journal = {Phys. Rev. A},
  volume = {92},
  issue = {2},
  pages = {023838},
  numpages = {6},
  year = {2015},
  month = {Aug},
  publisher = {American Physical Society},
  doi = {10.1103/PhysRevA.92.023838},
  url = {https://link.aps.org/doi/10.1103/PhysRevA.92.023838}
}

@article{ref23,
author = {Zhou, Yan-Hui and Liu, Tong and Zhang, Xing-Yuan and Wu, Qi-Cheng and Chen, Dong-Xu and Shi, Zhi-Cheng and Yang, Chui-Ping},
title = {Multimode Photon Blockade with a Reversed Design Method},
journal = {Advanced Quantum Technologies},
volume = {7},
number = {8},
pages = {2400089},
doi = {https://doi.org/10.1002/qute.202400089},
url = {https://advanced.onlinelibrary.wiley.com/doi/abs/10.1002/qute.202400089},
year = {2024}
}

@article{ref24,
  title={Multimode photon blockade},
  author={Chakram, Srivatsan and He, Kevin and Dixit, Akash V and Oriani, Andrew E and Naik, Ravi K and Leung, Nelson and Kwon, Hyeokshin and Ma, Wen-Long and Jiang, Liang and Schuster, David I},
  journal={Nature Physics},
  volume={18},
  number={8},
  pages={879--884},
  year={2022},
  publisher={Nature Publishing Group UK London},
  doi = {10.1038/s41567-022-01630-y},
  url = {https://doi.org/10.1038/s41567-022-01630-y},
}

@article{ref25,
author = {Zhang, Wei and Liu, Shutian and Zhang, Shou and Wang, Hong-Fu},
title = {Kerr-Nonlinearity Enhanced Photon Blockades via Driving a Δ-Type Atom},
journal = {Advanced Quantum Technologies},
volume = {6},
number = {12},
pages = {2300187},
doi = {https://doi.org/10.1002/qute.202300187},
url = {https://advanced.onlinelibrary.wiley.com/doi/abs/10.1002/qute.202300187},
year = {2023}
}

@article{ref26,
  title = {Exploring photon blockade in a two-photon Jaynes-Cummings model with atom and cavity drivings},
  author = {Li, Hai-Ji and Fan, Li-Bao and Ma, Shan and Liao, Jie-Qiao and Shu, Chuan-Cun},
  journal = {Phys. Rev. A},
  volume = {110},
  issue = {4},
  pages = {043707},
  numpages = {11},
  year = {2024},
  month = {Oct},
  publisher = {American Physical Society},
  doi = {10.1103/PhysRevA.110.043707},
  url = {https://link.aps.org/doi/10.1103/PhysRevA.110.043707}
}

@article{ref27,
doi = {10.1088/1674-1056/ac4cbc},
url = {https://doi.org/10.1088/1674-1056/ac4cbc},
year = {2022},
month = {jun},
publisher = {Chinese Physical Society and IOP Publishing Ltd},
volume = {31},
number = {7},
pages = {070304},
author = {Ding, Zhong and Zhang, Yong},
title = {Photon blockade in a cavity–atom optomechanical system},
journal = {Chinese Physics B}
}

@Article{ref28,
AUTHOR = {Li, Hong and Liu, Ming and Yang, Feng and Zhang, Siqi and Ruan, Shengping},
TITLE = {Phase-Controlled Tunable Unconventional Photon Blockade in a Single-Atom-Cavity System},
JOURNAL = {Micromachines},
VOLUME = {14},
pages = {2123},
YEAR = {2023},
NUMBER = {11},
ARTICLE-NUMBER = {2123},
URL = {https://www.mdpi.com/2072-666X/14/11/2123},
PubMedID = {38004980},
ISSN = {2072-666X},
DOI = {10.3390/mi14112123}

}

@article{ref29,
doi = {10.1088/1674-1056/ad8cbf},
url = {https://doi.org/10.1088/1674-1056/ad8cbf},
year = {2025},
month = {jan},
publisher = {Chinese Physical Society and IOP Publishing Ltd},
volume = {34},
number = {1},
pages = {014203},
author = {Luo, Ying and Zhang, Xinqin and Xiao, Yi and Xu, Jingping and Li, Haozhen and Yang, Yaping and Xia, Xiuwen},
title = {In-phase collective unconventional photon blockade and its stability in an asymmetrical cavity containing N bosonic atoms},
journal = {Chinese Physics B}
}

@article{ref30,
  title = {Nonreciprocal Photon Blockade},
  author = {Huang, Ran and Miranowicz, Adam and Liao, Jie-Qiao and Nori, Franco and Jing, Hui},
  journal = {Phys. Rev. Lett.},
  volume = {121},
  issue = {15},
  pages = {153601},
  numpages = {8},
  year = {2018},
  month = {Oct},
  publisher = {American Physical Society},
  doi = {10.1103/PhysRevLett.121.153601},
  url = {https://link.aps.org/doi/10.1103/PhysRevLett.121.153601}
}

@article{ref31,
  title = {Nonreciprocal unconventional photon blockade in a driven dissipative cavity with parametric amplification},
  author = {Shen, H. Z. and Wang, Qiu and Wang, Jiao and Yi, X. X.},
  journal = {Phys. Rev. A},
  volume = {101},
  issue = {1},
  pages = {013826},
  numpages = {13},
  year = {2020},
  month = {Jan},
  publisher = {American Physical Society},
  doi = {10.1103/PhysRevA.101.013826},
  url = {https://link.aps.org/doi/10.1103/PhysRevA.101.013826}
}

@article{ref32,
  title = {Nonreciprocal photon blockade and directional amplification in a spinning resonator coupled to a two-level atom},
  author = {Jing, Yu-Wei and Shi, Hai-Quan and Xu, Xun-Wei},
  journal = {Phys. Rev. A},
  volume = {104},
  issue = {3},
  pages = {033707},
  numpages = {9},
  year = {2021},
  month = {Sep},
  publisher = {American Physical Society},
  doi = {10.1103/PhysRevA.104.033707},
  url = {https://link.aps.org/doi/10.1103/PhysRevA.104.033707}
}

@article{ref33,
doi = {10.1088/1674-1056/ac523f},
url = {https://doi.org/10.1088/1674-1056/ac523f},
year = {2022},
month = {jun},
publisher = {Chinese Physical Society and IOP Publishing Ltd},
volume = {31},
number = {7},
pages = {074204},
author = {Zhang, Xinqin and Xia, Xiuwen and Xu, Jingping and Li, Haozhen and Fu, Zeyun and Yang, Yaping},
title = {Manipulation of nonreciprocal unconventional photon blockade in a cavity-driven system composed of an asymmetrical cavity and two atoms with weak dipole–dipole interaction},
journal = {Chinese Physics B}
}

@article{ref34,
author = {Luan, Tianze and Yang, Jiaxin and Wang, Jiao and Shen, Hongzhi and Zhou, Yanhui and Yi, Xuexi},
title = {Nonreciprocal unconventional photon blockade with spinning two-mode cavity coupled via χ(2) nonlinearities},
journal = {International Journal of Quantum Information},
volume = {21},
number = {05},
pages = {2350021},
year = {2023},
doi = {10.1142/S0219749923500211},
URL = {https://doi.org/10.1142/S0219749923500211},
}

@article{ref35,
author = {Shen, H. Z. and Luan, T. Z. and Zhou, Y. H. and Shi, Z. C. and Yi, X. X.},
title = {Nonreciprocal unconventional photon blockade in atom-cavity with χ(2) nonlinear medium},
journal = {International Journal of Quantum Information},
volume = {21},
number = {06},
pages = {2350029},
year = {2023},
doi = {10.1142/S0219749923500296},

URL = {  https://doi.org/10.1142/S0219749923500296}
}

@article{ref36,
doi = {10.1088/1674-1056/adb735},
url = {https://doi.org/10.1088/1674-1056/adb735},
year = {2025},
month = {apr},
publisher = {Chinese Physical Society and IOP Publishing Ltd},
volume = {34},
number = {5},
pages = {057202},
author = {Liu, Ming-Yue and Gong, Yuan and Chen, Jiaojiao and Wang, Yan-Wei and Xiong, Wei},
title = {Nonreciprocal microwave–optical entanglement in Kerr-modified cavity optomagnomechanics},
journal = {Chinese Physics B}
}

@article{ref37,
  title = {Nonreciprocal photon blockade induced by parametric amplification in an asymmetrical cavity},
  author = {Wu, Shao-Xiong and Gao, Xue-Chen and Cheng, Huan-Huan and Bai, Cheng-Hua},
  journal = {Phys. Rev. A},
  volume = {111},
  issue = {4},
  pages = {043714},
  numpages = {11},
  year = {2025},
  month = {Apr},
  publisher = {American Physical Society},
  doi = {10.1103/PhysRevA.111.043714},
  url = {https://link.aps.org/doi/10.1103/PhysRevA.111.043714}
}

@article{ref38,
author = {W. S. Xue and H. Z. Shen and X. X. Yi},
journal = {Opt. Lett.},
number = {16},
pages = {4424--4427},
publisher = {Optica Publishing Group},
title = {Nonreciprocal conventional photon blockade in driven dissipative atom-cavity},
volume = {45},
month = {Aug},
year = {2020},
url = {https://opg.optica.org/ol/abstract.cfm?URI=ol-45-16-4424},
doi = {10.1364/OL.398247},
}

@article{ref39,
  title = {Distinguishing photon blockade in a $\mathcal{PT}$-symmetric optomechanical system},
  author = {Wang, Dong-Yang and Bai, Cheng-Hua and Liu, Shutian and Zhang, Shou and Wang, Hong-Fu},
  journal = {Phys. Rev. A},
  volume = {99},
  issue = {4},
  pages = {043818},
  numpages = {10},
  year = {2019},
  month = {Apr},
  publisher = {American Physical Society},
  doi = {10.1103/PhysRevA.99.043818},
  url = {https://link.aps.org/doi/10.1103/PhysRevA.99.043818}
}

@article{ref40,
author = {Fan, Xiao-Hong and Zhang, Yi-Ning and Yu, Jun-Po and Liu, Ming-Yue and He, Wen-Di and Li, Hai-Chao and Xiong, Wei},
title = {Nonreciprocal Unconventional Photon Blockade with Kerr Magnons},
journal = {Advanced Quantum Technologies},
volume = {7},
number = {8},
pages = {2400043},
doi = {https://doi.org/10.1002/qute.202400043},
url = {https://advanced.onlinelibrary.wiley.com/doi/abs/10.1002/qute.202400043},
year = {2024}
}

@article{ref41,
  title = {Nonreciprocal photon-phonon entanglement in Kerr-modified spinning cavity magnomechanics},
  author = {Chen, Jiaojiao and Fan, Xiao-Gang and Xiong, Wei and Wang, Dong and Ye, Liu},
  journal = {Phys. Rev. A},
  volume = {109},
  issue = {4},
  pages = {043512},
  numpages = {9},
  year = {2024},
  month = {Apr},
  publisher = {American Physical Society},
  doi = {10.1103/PhysRevA.109.043512},
  url = {https://link.aps.org/doi/10.1103/PhysRevA.109.043512}
}

@article{ref42,
  title = {Nonlinear dissipation-induced photon blockade},
  author = {Su, Xin and Tang, Jiang-Shan and Xia, Keyu},
  journal = {Phys. Rev. A},
  volume = {106},
  issue = {6},
  pages = {063707},
  numpages = {13},
  year = {2022},
  month = {Dec},
  publisher = {American Physical Society},
  doi = {10.1103/PhysRevA.106.063707},
  url = {https://link.aps.org/doi/10.1103/PhysRevA.106.063707}
}

@article{ref43,
author = {Xie, Hong and He, Le-Wei and Shang, Xiao and Lin, Xiu-Min},
title = {Photon Blockade in Cavity Optomechanics Via Parametric Amplification},
journal = {Advanced Quantum Technologies},
volume = {7},
number = {9},
pages = {2400065},
doi = {https://doi.org/10.1002/qute.202400065},
url = {https://advanced.onlinelibrary.wiley.com/doi/abs/10.1002/qute.202400065},
year = {2024}
}

@article{ref44,
doi = {10.1088/1464-4266/1/4/312},
url = {https://doi.org/10.1088/1464-4266/1/4/312},
year = {1999},
month = {aug},
publisher = {},
volume = {1},
number = {4},
pages = {424},
author = {Tan, S M},
title = {A computational toolbox for quantum and atomic optics},
journal = {Journal of Optics B: Quantum and Semiclassical Optics}
}

@misc{ref45,
  author = {Tan, S M},
  title = {Quantum Optics Toolbox for MATLAB},
  year = {2012-12-21},
  howpublished = {\url{https://github.com/jevonlongdell/qotoolbox}},
}

@article{ref46,
  title = {Unconventional Photon Blockade Under the Combined Effects of Second-Order Nonlinearity and Two-Photon Absorption Environment},
  author = {Zhang, Zhi Qiang },
  journal = {Laser \& Optoelectronics Progress},
  volume = {62},
  issue = {7},
  pages = {0719001},
  year = {2025},
  doi = {http://dx.doi.org/10.3788/LOP242095},
  url = {http://dx.doi.org/10.3788/LOP242095}
}

@article{ref47,
  title = {Simultaneous nonreciprocal photon blockade via directional parametric amplification},
  author = {Zhang, Wei and Hou, Rui and Wang, Tie and Liu, Shutian and Zhang, Shou and Wang, Hong-Fu},
  journal = {Phys. Rev. A},
  volume = {110},
  issue = {2},
  pages = {023723},
  numpages = {9},
  year = {2024},
  month = {Aug},
  publisher = {American Physical Society},
  doi = {10.1103/PhysRevA.110.023723},
  url = {https://link.aps.org/doi/10.1103/PhysRevA.110.023723}
}

@article{ref48,
  title = {Giant Enhancement of Unconventional Photon Blockade in a Dimer Chain},
  author = {Wang, You and Verstraelen, W. and Zhang, Baile and Liew, Timothy C. H. and Chong, Y. D.},
  journal = {Phys. Rev. Lett.},
  volume = {127},
  issue = {24},
  pages = {240402},
  numpages = {7},
  year = {2021},
  month = {Dec},
  publisher = {American Physical Society},
  doi = {10.1103/PhysRevLett.127.240402},
  url = {https://link.aps.org/doi/10.1103/PhysRevLett.127.240402}
}

@article{ref49,
  title = {Universal Photon Blockade},
  author = {Zhou, Yan-Hui and Liu, Tong and Su, Qi-Ping and Zhang, Xing-Yuan and Wu, Qi-Cheng and Chen, Dong-Xu and Shi, Zhi-Cheng and Shen, H. Z. and Yang, Chui-Ping},
  journal = {Phys. Rev. Lett.},
  volume = {134},
  issue = {18},
  pages = {183601},
  numpages = {7},
  year = {2025},
  month = {May},
  publisher = {American Physical Society},
  doi = {10.1103/PhysRevLett.134.183601},
  url = {https://link.aps.org/doi/10.1103/PhysRevLett.134.183601}
}

\end{document}